\documentclass[10pt,letterpaper]{article}
\usepackage[margin=0.85in]{geometry}
\usepackage[hyphens]{url}
\usepackage{graphicx}
\usepackage{dblfloatfix}
\usepackage{float}
\usepackage{natbib}
\usepackage{caption}
\usepackage{amsmath}
\usepackage{amssymb}
\usepackage{adjustbox}
\usepackage{booktabs}
\usepackage{xcolor}
\usepackage{enumitem}
\usepackage{authblk}
\usepackage{natbib}
\setcitestyle{authoryear,round,semicolon,aysep={}}
\let\cite\citep

\usepackage[colorlinks=true,
            citecolor=black,
            linkcolor=black,
            urlcolor=black]{hyperref}

\title{What Must LLM Agents Communicate?\\ From Information Loss to Communication Failures}
\author{
    Yi-Yun Chu\textsuperscript{\rm 1},
    Xiaobin Shen\textsuperscript{\rm 2}
}
\affil{
    \textsuperscript{\rm 1}McGill University,
    \textsuperscript{\rm 2}Carnegie Mellon University\\
    yi-yun.chu@mcgill.ca,
    xiaobins@andrew.cmu.edu
}

\newcommand{\logit}{\operatorname{logit}}

\begin{document}
\maketitle

\begin{abstract}
When evidence is distributed across LLM agents, a team may collectively possess all the information needed for a decision yet fail to use it effectively. What must agents communicate for accurate joint predictions, and when does interaction help? We study three requirements for effective communication: sufficiency of the information conveyed, selection of relevant evidence, and utilization of received information. Drawing on information theory and communication complexity, we develop a framework that combines theoretical benchmarks with controlled experiments and message-level analyses. We find that local probability reports can discard essential information, with theoretical information loss closely matching observed performance penalties. Query-first interaction improves predictions when relevant evidence depends on a private query, but compact one-shot representations can eliminate this advantage. Even when sufficient information is communicated, agents may fail to select or use it correctly. Together, these findings distinguish fundamental communication limits from failures in how LLM agents communicate and use information.
\end{abstract}

\section{Introduction}
LLM agents increasingly collaborate on tasks where different agents have access to different pieces of information. Even when a team collectively possesses everything needed for a decision, its performance depends on what its members communicate. Consider an agent that knows which connector a customer ordered and another that knows which connector was delivered. Neither can determine whether the order was fulfilled correctly on its own, but exchanging the two connector types resolves the question. Exchanging only their separate estimates of whether the connectors match, however, may reveal nothing: if the required and delivered types are independently drawn from two equally likely possibilities, both agents correctly report a 50\% probability of a match regardless of what they observe. This illustrates a central challenge of collaboration under distributed information: what must agents communicate to make accurate joint predictions, and when is it useful to ask for information before responding?

Prior research has examined communication among LLM agents, from debate and aggregation of independently generated answers~\cite{du2024debate,smit2024mad, choi2025debate, wynn2025talk} to collaboration in which agents hold different pieces of information~\cite{li2026hiddenbench, li2026infodelphi}. These studies show that communication can improve performance, but can also introduce errors. Yet final performance alone does not explain why a communication strategy succeeds or fails. Poor predictions might arise because messages discard essential information, relevant evidence is never sent, or the receiving agent fails to use it correctly. Distinguishing these possibilities helps explain when communication is effective and why it sometimes fails.

\begin{figure*}[htbp]
\centering
\includegraphics[width=1\textwidth]{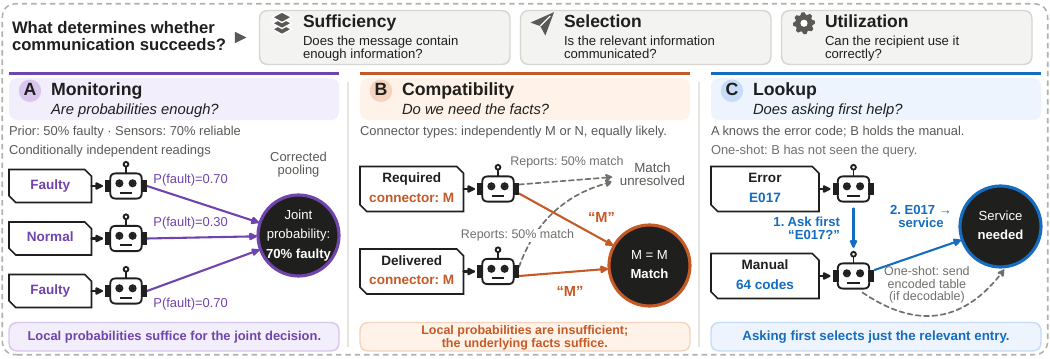}
\vspace{-0.5em}
\caption{\textbf{Communication under distributed information.}
Agents hold different pieces of evidence needed for a joint decision.
(A) \emph{Monitoring:} With conditionally independent sensor readings, local probability reports suffice to recover the joint posterior.
(B) \emph{Compatibility:} Each agent knows one connector type and assigns a 50\% probability to a match. Exchanging these probabilities leaves the outcome unresolved, whereas exchanging the types determines it.
(C) \emph{Lookup:} Asking for an error code allows the manual holder to send the relevant entry, whereas one-shot communication must support an unknown query.
Together, the examples illustrate information sufficiency, evidence selection, and message utilization.}
\label{fig:framework}
\end{figure*}

We distinguish three requirements for effective communication: \emph{sufficiency}, whether a communication format preserves the information needed for the joint prediction; \emph{selection}, whether agents can identify and transmit relevant evidence; and \emph{utilization}, whether recipients correctly use the information they receive (Figure~\ref{fig:framework}). Bayesian aggregation and communication complexity provide theoretical benchmarks for what different communication strategies can achieve~\cite{bordley1982multiplicative, dietrich2010bayesian, kremer1999oneround, kushilevitz1997communication}. We evaluate LLM agents against these benchmarks using controlled tasks with exact Bayesian references and audits of the messages exchanged, distinguishing limitations of communication strategies from errors in their execution.

\newpage
Our contributions are threefold:
\begin{itemize}[noitemsep,topsep=0pt, leftmargin=1em]
    \item We develop a diagnostic framework combining theoretical benchmarks and message-level audits to evaluate information sufficiency, evidence selection, and recipient utilization.
    \item We characterize when local probability reports suffice for joint prediction, quantify their intrinsic information loss, and connect these limits to observed LLM performance.
    \item We show how the value of query-first interaction depends on query visibility and message representation, and how recipient reasoning affects the use of complete messages.
\end{itemize}

\section{Framework}
\label{sec:framework}

We consider binary prediction tasks with outcome $Y\in\{0,1\}$. Each agent $i$ observes private evidence $U_i$ and information $S$ shared by all agents. A final decision-maker predicts the probability of $Y=1$ using the information communicated by the agents.

We evaluate predictions using logarithmic loss, which penalizes assigning low probability to the outcome that occurs. Given shared information $S$ and additional evidence $Z$, the optimal prediction is $\Pr(Y=1\mid S,Z)$. Its minimum expected log loss is $R^*(Z)=H(Y\mid S,Z)$, where $H$ denotes conditional entropy~\cite{gneiting2007proper,cover2006elements}. By considering different information sets $Z$, such as communicated probability reports or all private evidence $U=(U_1,\ldots,U_n)$, we can compare the best predictive performance achievable from each.

\noindent\textbf{Sufficiency.}
Sufficiency concerns whether a message representation contains enough information for the joint prediction. We examine whether local probability reports can replace underlying observations without losing decision-relevant information. Let $P_i^*=\Pr(Y=1\mid S,U_i)$ denote the exact Bayesian probability implied by agent $i$'s evidence. An LLM's actual report, $\widehat P_i$, may differ because of probability-estimation or reporting errors. To isolate limitations of the communication format from such sender-side errors, we consider exact local reports.

For the vector $P^*=(P_1^*,\ldots,P_n^*)$, we define the \emph{reporting loss} as
\begin{equation}
\Delta_P=R^*(P^*)-R^*(U)=I(Y;U\mid S,P^*),
\label{eq:reporting}
\end{equation}
where $I$ denotes conditional mutual information. The difference between the minimum expected log loss using local probability reports, $R^*(P^*)$, and that using all underlying evidence, $R^*(U)$, measures information lost even with perfectly accurate reports (Appendix~\ref{app:theory}). Thus, $\Delta_P=0$ means the reports are sufficient for the joint prediction, whereas $\Delta_P>0$ indicates information loss inherent to the representation.

When private observations are conditionally independent given $Y$ and $S$, the joint posterior can be recovered from local probabilities by correcting for shared information counted multiple times~\cite{bordley1982multiplicative,dietrich2010bayesian}. We call this \emph{corrected Bayesian pooling}, illustrated by monitoring in Figure~\ref{fig:framework}A, where $\Delta_P=0$. In contrast, in Figure~\ref{fig:framework}B, each agent's probability of a connector match remains one-half regardless of its observation, while the connector types together determine the outcome. Hence $R^*(P^*)=\log 2$, $R^*(U)=0$, and $\Delta_P=\log 2$ nats.

\noindent\textbf{Selection.}
Selection concerns whether the sender can identify and communicate the evidence relevant to the joint prediction. In the lookup task (Figure~\ref{fig:framework}C), one agent holds a binary table $X=(X_1,\ldots,X_K)$, while another knows the requested index $J$, with target $Y=X_J$. Without knowing $J$, the table holder must communicate enough information to answer an unknown query. If the index holder speaks first, the table holder can instead send only the requested entry.

Classical communication complexity characterizes this distinction: answering an unknown index reliably requires $\Omega(K)$ bits in one-way communication, whereas revealing the index first requires $\lceil\log_2 K\rceil$ bits for the query and one bit for the reply~\cite{kremer1999oneround,kushilevitz1997communication}. This motivates a potential advantage of interaction, but does not guarantee that LLMs will realize it under token budgets.

To measure this advantage, we compare one-shot and query/reply communication when the query is initially private or public. Let $R^v_{\pi}(b)$ denote the expected log loss of protocol $\pi$ under query visibility $v$ and delivery budget $b$. Using matched task instances, we define
\begin{equation}
D(b)=\big(R^{\mathrm{priv}}_{\mathrm{one}}(b)-R^{\mathrm{priv}}_{\mathrm{ask}}(b)\big)
-\big(R^{\mathrm{pub}}_{\mathrm{one}}(b)-R^{\mathrm{pub}}_{\mathrm{ask}}(b)\big).
\label{eq:dd}
\end{equation}
A positive $D(b)$ indicates that asking first provides a larger improvement when the query is private. Unlike $\Delta_P$, this compares the observed performance of complete protocols, including their message representations and model execution.

\noindent\textbf{Utilization.}
Utilization concerns whether the recipient correctly uses the information it receives. Even when a message contains sufficient evidence, an LLM may misread an encoding or make an incorrect inference. We investigate these failures by auditing which facts reach the final reader and comparing its predictions with those of a full-information reader, the same model given all private evidence. Both are evaluated against the exact Bayesian reference.

When sufficient information reaches the reader, a performance shortfall relative to the Bayesian reference cannot be explained by missing evidence alone. Conversely, missing relevant facts may indicate selection failures, although our recorded sufficient fact sets are not exhaustive. These comparisons help distinguish missing information from errors in using the information available.

\section{Experimental Setup}
\label{sec:setup}

We generate synthetic cases from specified probability distributions and decision rules so that exact Bayesian predictions serve as evaluation references~\citep{weston2016towards,saparov2023language}. Figure~\ref{fig:framework} illustrates the three core tasks.

\noindent\textbf{Tasks.}
(i) \textbf{\textit{Monitoring}} is the case in which probability reports suffice~\citep{genest1986combining}: three agents each observe a private, 70\%-reliable sensor reading about a machine's state, alongside a shared inspection. We enumerate 32 observation patterns across two inspection reliabilities, weighted by their exact probabilities.
(ii) \textbf{\textit{Compatibility}} combines equality and threshold checks~\citep{kushilevitz1997communication}: two agents separately observe an order's requirements and a product's specifications, including connector type, capacity, and temperature, along with irrelevant attributes. Of 64 possible order--product combinations, four are used for development and the remaining 60 for evaluation.
(iii) \textbf{\textit{Lookup}} is the INDEX function~\citep{kremer1999oneround}, a multi-agent form of key-value retrieval~\citep{liu2024lost}: one agent knows an error code, while another holds a balanced manual of $K\in\{16,64\}$ binary service decisions. We generate 50 manuals per size, each paired with a query evaluated under both private and public visibility, keeping the underlying task identical.

\noindent\textbf{Reporting-loss sweep.}
We add a sensor--parity task to test whether $\Delta_P$ predicts the observed pooling penalty. Each of two agents sees a noisy sensor reading and one code bit. The bits provide additional evidence only when combined~\citep{williams2010nonnegative}. Nine settings vary $\Delta_P$, including three at a fixed full-information Bayes risk of 0.300 nats. We compare pooling with complete factual collection without a delivery cap (Appendix~\ref{app:tasks}).

\noindent\textbf{Protocols.}
We compare probability reporting with factual communication. For probability reporting, LLM agents produce local forecasts, which are combined deterministically using the arithmetic mean, summed log-odds, or corrected Bayesian pooling; the latter also uses a public-only forecast. For factual communication, agents select fact identifiers, and a relay copies the corresponding values into indexed messages (e.g., \texttt{E017=1}). This allows agents to choose what to communicate without altering the selected facts.

\emph{Fixed collection} deterministically selects eligible facts in a prescribed order within the budget; in lookup, it sends available rows when the query is private and the requested row when the query is public. \emph{Independent selection} and \emph{sequential discussion} use LLMs to select facts, with later speakers in discussion seeing earlier messages. Under tight budgets, discussion is primarily sequential factual exchange rather than unrestricted debate. In \emph{query/reply}, the query holder speaks first, allowing the table holder to select the requested entry. We also test deterministic \emph{hex encoding}, which compresses the lookup table in a shared row order and sends the longest prefix that fits when the query is private. Factual and encoded messages are passed to a final LLM reader; probability reports are combined by the specified aggregation rule (Appendix~\ref{app:protocols}).

\noindent\textbf{Models and budgets.}
We primarily evaluate quantized Qwen3.5-9B and Qwen3.5-27B, using the same model for all agents within a condition. The core experiments use delivery budgets of 64, 256, and 1,024 native tokens, counting each message delivered to a recipient and withholding oversized messages. Separately, LLM calls have a 4,096-token reasoning allowance, including for the final reader. Sequential protocols divide the delivery budget among senders and provide guidance on the number of facts to transmit. The graded sensor--parity sweep has no delivery cap. Gemma-4-E4B-it provides a cross-family check on monitoring, compatibility, and 50 lookup worlds at 64 delivered tokens. A Qwen 27B follow-up tests hex-encoded tables of up to 1,024 entries on 25 worlds per size. For the 1,024-entry tables, we rerun the reader with identical prompts and a 32,768-token reasoning allowance. Main experiments use one temperature-zero generation per model call; details are in Appendix~\ref{app:execution}.

\noindent\textbf{Outcomes.}
Our primary outcome is expected log loss, evaluated against the exact full-information posterior $q=\Pr(Y=1\mid S,U)$, integrating over any remaining label uncertainty. For compatibility and lookup, where complete evidence determines a binary answer, we also report prediction accuracy. Invalid final outputs are retained, receiving a fixed loss of $-\log 10^{-6}\approx13.8$ nats and zero accuracy.

The message audit uses the full-information reader control from Section~\ref{sec:framework}. Our primary interaction comparison is $D(64)$ for Qwen 27B, using fixed collection as the one-shot baseline. Its 95\% confidence interval is computed by bootstrapping 100 lookup worlds while preserving private/public pairs and stratifying by table size; the Gemma and compact-encoding comparisons are exploratory.

\section{Results}
\label{sec:results}

Figure~\ref{fig:results} summarizes the main findings; detailed results appear in Appendices~\ref{app:results} and~\ref{app:robust}.

\begin{figure*}[htbp]
\centering
\includegraphics[width=0.97\textwidth]{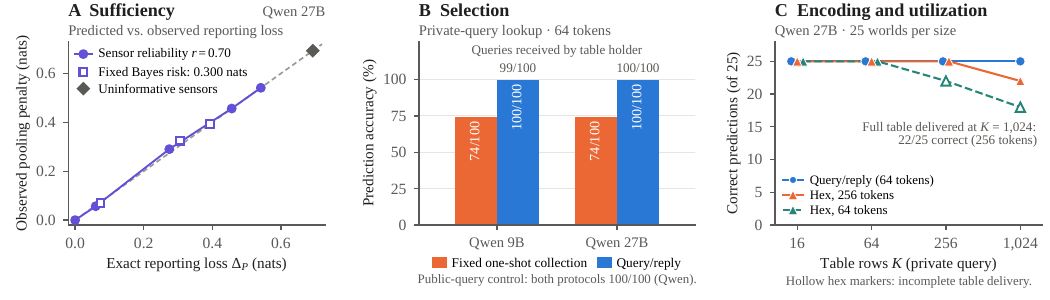}
\vspace{-0.5em}
\caption{\textbf{Main results.}
(A) \emph{Sufficiency:} For Qwen 27B, the observed pooling penalty (corrected pooling minus full-information reading) closely matches the theoretical reporting loss $\Delta_P$ across nine graded settings; the dashed line indicates equality.
(B) \emph{Selection:} Query/reply outperforms fixed one-shot collection on private-query lookup at 64 delivered tokens for both Qwen models. Annotations count queries received by the table holder; with public queries, both protocols achieve 100/100 accuracy.
(C) \emph{Encoding and utilization:} Query/reply uses a 64-token delivery budget, while hex-encoded one-shot messages are evaluated at 64 and 256 delivered tokens. All results use a 4,096-token reasoning allowance for the final reader. Hollow hex markers indicate incomplete table delivery. At $K=1024$, the complete table fits within 256 delivered tokens, yet the reader answers only 22/25 cases correctly.
}\vspace{-1em}
\label{fig:results}
\end{figure*}

\subsection{Sufficiency: what probability reports preserve}

We first examine when local probability reports contain enough information for the joint prediction. In monitoring, where private readings are conditionally independent given the machine's state and shared evidence, corrected Bayesian pooling recovers the exact joint posterior across all 32 observation patterns for both Qwen models. In compatibility, however, corrected pooling yields only 44 correct predictions out of 60 held-out cases for all three models, compared with 60/60 when a fixed collection rule forwards the underlying specifications to an LLM reader. Both Qwen models report exact local probabilities, so their performance gap reflects information lost through probability reporting rather than estimation errors.

We next test whether the theoretical reporting loss $\Delta_P$ predicts the observed penalty. In the graded sensor--parity task, the expected log-loss difference between corrected pooling and a full-information LLM reader differs from $\Delta_P$ by at most 0.020 nats across nine settings for Qwen 27B (Figure~\ref{fig:results}A), including settings with identical full-information Bayes risk. Qwen 9B shows larger deviations associated with local probability-estimation and full-information reader errors (Appendix~\ref{app:graded-9b}).

\subsection{Selection: when asking first helps}

We compare one-shot lookup, where the table holder sends all entries without knowing the query, with query/reply, where the requested code is communicated first. At 64 delivered tokens, query/reply answers all 100 private-query cases correctly for both Qwen models, compared with 74/100 for fixed one-shot collection (Figure~\ref{fig:results}B). When the query is public, both protocols achieve 100/100. For Qwen 27B, the private-versus-public contrast is $D(64)=0.484$ nats (95\% CI [0.421, 0.548]).

The advantage also depends on message representation. For $K\le64$, the complete hex-encoded table and separately transmitted query fit within 64 tokens, and both hex encoding and query/reply answer all 50 sampled cases correctly (Figure~\ref{fig:results}C). Thus, asking first provides an advantage under the indexed-message format, but not when a sufficiently compact one-shot representation is available.

\subsection{Utilization: when complete messages are not enough}
\label{sec:utilization}

We examine whether the final reader can correctly use information that has been fully delivered, separating communication limits from reasoning limits. In lookup with $K=1024$, the complete hex-encoded table fits within a 256-token delivery budget, yet Qwen 27B answers only 22 of 25 cases correctly with a 4,096-token reader reasoning allowance, compared with 25/25 for query/reply using 64 delivered tokens (Figure~\ref{fig:results}C). Thus, increasing the communication budget sufficiently to deliver the entire table does not guarantee that the reader can use it correctly. A matched rerun with identical reader prompts and a 32,768-token reasoning allowance achieves 25/25 (Appendix~\ref{sec:compact}).

Model execution also varies across families and tasks. Gemma omits the query in 32 of 50 lookup exchanges, with query/reply answering 35/50 cases compared with 37/50 for fixed collection. It also produces three invalid monitoring predictions despite receiving all private readings (Appendix~\ref{sec:crossfamily}).

\section{Discussion}

Our results distinguish information lost through communication from failures in selecting and using evidence. Theoretical reporting loss identifies intrinsic limits of probability reports, while message audits and full-information controls reveal model execution errors that final performance alone cannot explain.

Interaction is not always necessary: compact one-shot messages can match query/reply, while complete messages can still be difficult for LLMs to use. Fewer delivered tokens can also require more inference. Communication strategies should therefore be evaluated jointly for information sufficiency, model execution, and computational cost.

Our controlled synthetic tasks and structured messages enable precise evaluation but leave open how these findings generalize. Natural extensions include distributed-information benchmarks such as HiddenBench~\cite{li2026hiddenbench}, broader model families, and comparisons under matched inference costs.

\bibliographystyle{plainnat}
\bibliography{references}

\clearpage
\renewcommand{\thesection}{\Alph{section}}
\renewcommand{\thesubsection}{\Alph{section}\arabic{subsection}}
\counterwithin{table}{section}
\counterwithin{figure}{section}
\counterwithin{equation}{section}
\renewcommand{\thetable}{\Alph{section}\arabic{table}}
\renewcommand{\thefigure}{\Alph{section}\arabic{figure}}
\appendix

\section{Related Work}
\label{app:related}

\paragraph{Debate and deliberation among LLM agents.}
In multi-agent debate, model instances exchange answers and reasoning over several rounds, which can improve factuality and reasoning~\cite{du2024debate}. Later work questions where the gains come from. Debate protocols do not reliably beat self-consistency or ensembling unless carefully tuned, and they are sensitive to how strongly agents are prompted to agree~\cite{smit2024mad}. Modeling debate as a stochastic process shows that it forms a martingale over agents' beliefs, and majority voting accounts for most of its reported gains~\cite{choi2025debate}. Exchange can also hurt: agents abandon correct answers for flawed peer reasoning~\cite{wynn2025talk}, similar models fall into echo chambers~\cite{estornell2024debate}, and errors are correlated across LLMs, which limits what independent votes add~\cite{kim2025correlated}. In these settings every agent sees the same question and evidence, so deliberation mainly reweights or revises answers. We study the complementary case in which evidence is distributed, so some exchange is needed to move information at all, and ask which representation of it must
move and when.

\paragraph{Distributed information in human and LLM groups.}
Hidden-profile experiments show that human groups discuss shared information more than unshared information and therefore miss decisions that require pooling it~\cite{stasser1985pooling, lu2012hidden}. HiddenBench brings the paradigm to LLM
teams: across 15 models, groups with distributed information reach 30.1\% accuracy against 80.7\% for a single agent holding all of it, and the authors evaluate interventions that disclose all facts or structure exchange before a decision~\cite{li2026hiddenbench}. InfoDelphi combines evidence routing, rationale exchange, and confidence-weighted aggregation for forecasting under information asymmetry~\cite{li2026infodelphi}, and CrossProfile varies fact ownership and usefulness to study how selection and attribution shape group decisions~\cite{imel2026crossprofile}. These studies establish that pooling fails and that structured disclosure helps. Because our tasks have exact Bayes references, we can ask a different question: whether a given channel (probabilities, facts, or a query-first exchange) can carry the needed information at all, and when it can, whether agents select and use it.

\paragraph{Aggregating probabilities.}
Opinion pooling studies how to combine probability judgments~\cite{genest1986combining}. When experts' information is conditionally independent, Bayesian aggregation takes a multiplicative form~\cite{bordley1982multiplicative}, and communicating prior and posterior beliefs suffices without disclosing the underlying evidence~\cite{dietrich2010bayesian}; when information overlaps arbitrarily, the correct rule is more complex. Logit-scale aggregation with extremizing improves forecasts from partially informed forecasters~\cite{satopaa2014combining}, and proper scoring rules~\cite{gneiting2007proper} evaluate the resulting probabilities. Our corrected pooling rule is the conditionally independent case, and our compatibility and graded tasks are built so that its assumptions fail by a known amount.

\paragraph{Eliciting probabilities from LLMs.}
Larger language models are fairly well calibrated on multiple-choice and true/false questions and can estimate the probability that their own answers are correct~\cite{kadavath2022know}. For models fine-tuned with human feedback, verbalized confidence is often better calibrated than token probabilities~\cite{tian2023justask}, although it tends toward overconfidence and benefits from consistency across sampled responses~\cite{xiong2024uncertainty}. Simple aggregation of forecasts from a dozen LLMs can rival a human crowd~\cite{schoenegger2024silicon}. These studies evaluate calibration against realized outcomes. Pooling needs something stricter: each agent must report the exact conditional probability implied by its evidence and the stated distribution, because the aggregation rule assumes truthful posteriors. Our references check this directly. Both Qwen models report exact local posteriors in monitoring, whereas Gemma there and Qwen 9B in the graded task do not.

\paragraph{Communication complexity and learned communication.}
Two-party communication complexity counts the bits needed to compute a function of distributed inputs~\cite{yao1979complexity,kushilevitz1997communication}. INDEX separates one-way from interactive protocols~\cite{kremer1999oneround}, and pointer chasing shows how the number of rounds matters~\cite{nisan1993rounds}. Multi-agent reinforcement learning learns what to communicate~\cite{sukhbaatar2016commnet,foerster2016learning}, to whom~\cite{das2019tarmac}, and when~\cite{singh2019ic3net,liu2020when2com}. For LLM systems, recent work reduces communication cost by pruning redundant messages~\cite{zhang2025agentprune}, training agents for token efficiency~\cite{chen2025optima}, or exchanging embeddings instead of text~\cite{pham2024cipher}. These methods optimize communication for task performance. We instead fix a task's information structure and measure what a message must contain and when it must be sent, separating the limits of a channel from the failures of the agents using it.

\section{Theory Details}
\label{app:theory}

\subsection{Oracle identities}
For any available information $Z$, minimizing conditional expected log loss
over probability forecasts yields the conditional entropy $H(Y\mid S,Z)$.
Since $P^*$ is a deterministic function of $(S,U)$,
\begin{align*}
&H(Y\mid S,P^*)-H(Y\mid S,U)\\
&\quad=H(Y\mid S,P^*)-H(Y\mid S,U,P^*)\\
&\quad=I(Y;U\mid S,P^*).
\end{align*}
Equivalently, this difference is the expected KL divergence between
$p(Y\mid S,U)$ and $p(Y\mid S,P^*)$. The identity concerns truthful local
posteriors under the stated distribution; the empirical pooling rules use
elicited probabilities and specified numerical aggregators.

\subsection{Corrected pooling}
Conditional independence of the private signals given $Y$ and $S$ is sufficient for probability reports to determine the joint prediction. When the relevant probabilities lie strictly between zero and one, Bayes' rule gives
\begin{align}
\logit\Pr(Y=1\mid S,U)
&=\sum_{i=1}^n\logit P_i^* \nonumber\\
&\quad -(n-1)\logit\Pr(Y=1\mid S).
\label{eq:pool}
\end{align}
The subtraction corrects for the public information counted in every local posterior. This is the multiplicative aggregation rule of \citet{bordley1982multiplicative} and the conditionally independent-information case of \citet{dietrich2010bayesian}. Our monitoring task satisfies these assumptions, so corrected pooling can recover the joint posterior from local probability reports.

For Equation~\ref{eq:pool}, let $O_S$ be the public posterior odds and
$L_i=p(U_i\mid Y=1,S)/p(U_i\mid Y=0,S)$. Local posterior odds are
$O_i=O_SL_i$, and conditional independence gives joint odds
$O_S\prod_iL_i=(\prod_iO_i)/O_S^{n-1}$. Taking logarithms yields the corrected
pooling rule. In simple matching, $P_1^*=P_2^*=1/2$,
$H(Y\mid S,P^*)=\log2$, and $H(Y\mid S,U)=0$.

By contrast, consider two independent, uniformly distributed binary interfaces $U_1$ and $U_2$, with compatibility defined by $Y=\mathbf{1}\{U_1=U_2\}$. Each agent's truthful local posterior is always $1/2$, regardless of its observed interface. The reports therefore reveal nothing about whether the interfaces match, whereas the two underlying values determine $Y$ exactly. Consequently, $\Delta_P=\log 2$ nats. This example illustrates why exchanging underlying facts can be necessary even when every agent reports its local probability perfectly.

\subsection{The value of interaction}
Let $R_{\pi,v}(b)$ denote the expected log loss of protocol $\pi$, where $v\in\{\mathrm{private},\mathrm{public}\}$ specifies whether the query is initially visible to the table holder, and $b$ is the delivery-token budget. Define
\begin{align}
G_v(b)
&=R_{\mathrm{one\text{-}shot},v}(b)
-R_{\mathrm{query},v}(b),\nonumber\\
D(b)
&=G_{\mathrm{private}}(b)
-G_{\mathrm{public}}(b),
\label{eq:adaptation}
\end{align}
which is Equation~\ref{eq:dd} in the main text.
Here, $G_v(b)>0$ indicates that query/reply outperforms one-shot communication under visibility condition $v$. The contrast $D(b)$ measures whether this advantage is larger when the query is initially private. By evaluating both visibility conditions on the same $X$, $J$, and $Y$, we isolate the additional benefit associated with query-dependent selection. Unlike $\Delta_P$, these quantities compare complete empirical protocols, including their message representations and inference calls.

This distinction has a classical counterpart in communication complexity. Distributed lookup corresponds to the INDEX function: if only the table holder may speak, reliably answering an unknown query requires $\Omega(K)$ bits of communication~\cite{kremer1999oneround,kushilevitz1997communication}. If the index holder speaks first, $\lceil\log_2 K\rceil$ bits for the query and one bit for the response suffice. This motivates the potential advantage of interaction, but does not guarantee that it will improve LLM performance. Actual performance also depends on how efficiently information is encoded, whether agents select and transmit the necessary facts, and whether the receiving LLM can use them correctly. Our experiments therefore compare communication protocols under explicit token budgets and examine these execution failures separately.

\section{Tasks, Protocols, and Implementation}
\label{app:tasks}

\subsection{Controlled task suite}
\paragraph{Monitoring.}
A machine is faulty with prior probability $1/2$. A shared inspection and three private instruments give binary readings that are independent conditional on the machine's state. Each private instrument is correct with probability $0.70$; the shared inspection has reliability $0.50$ or $0.75$. We enumerate all 16 reading patterns for each shared reliability, giving 32 cases weighted by their exact probabilities.

\paragraph{Compatibility.}
The simple task contains the four combinations of two equally likely interfaces. The detailed task adds capacity and operating-temperature requirements: the order and the product each specify one of two interfaces, capacities, and temperatures. A delivery is compatible when its interface matches and its capacity and temperature ratings meet the order's requirements. All $8\times8=64$ combinations are enumerated, each with an irrelevant folder color and batch tag. Four combinations were used in development; the other 60 form the held-out evaluation set.

\paragraph{Lookup.}
A maintenance manual maps each error code to a binary service decision, with exactly half of its $K$ entries requiring service. We sample 50 balanced manuals at each of $K\in\{16,64\}$, with an independent uniform query for each, none overlapping the development manuals. Each world appears twice: the query is private to the query holder, or public to all participants. The compact-encoding, reasoning, and Gemma experiments use a fixed, label-blind subset of 25 manuals per size, and the compact study adds 25 manuals at each of $K\in\{256,1024\}$.

\paragraph{Graded sensor--parity task.}
Two agents each hold a noisy sensor reading and one bit of a two-bit code whose parity is a further noisy signal about the same state; neither code bit alone is informative, so local posteriors cannot carry the parity clue. Nine settings, each enumerating all 32 observations with exact weights, vary $\Delta_P$: five vary parity noise at sensor reliability 0.70, three hold the full-information entropy at 0.300 nats while varying sensor reliability, and one has uninformative sensors and noiseless parity (Appendix~\ref{app:sensitivity}).

Monitoring, compatibility, and lookup form the core suite of 300 case versions from 200 worlds. Exact posteriors, labels, and sufficient fact sets are retained by the evaluator; agents receive their allocated evidence and the public sampling rules.

\subsection{Communication protocols}
\label{app:protocols}
All evidence protocols end with a final reader that sees the public task information and the delivered messages, with core instructions shared across conditions; each condition uses agents of one model. Table~\ref{tab:protocols} lists message flow and inference calls.

\paragraph{Probability pooling.}
Agents report binary probability vectors, combined by their arithmetic mean, by summed log odds, or by the corrected rule in Equation~\ref{eq:pool} with an elicited public-only probability. Pooling needs $n$ or $n+1$ forecasts.

\paragraph{Factual collection.}
A relay copies the values of selected private fact IDs into indexed messages such as \texttt{E017=1}, so a sender chooses which facts to send but cannot alter them. \emph{Independent selection} uses an LLM sender per agent. \emph{Fixed collection} is a deterministic reference policy that omits the explicitly irrelevant fields and includes eligible facts in order while they fit; for private-query lookup it sends a prefix of the manual and reserves space for the query. It needs a single reader call. Complete disclosure supplies all private facts where its cost fits the budget.

\paragraph{Sequential exchange.}
In query/reply, the query holder sends first, the table holder responds with selected facts, and the reader decides (three calls). Sequential discussion uses the same order in lookup and a fixed agent order elsewhere, with factual attachments and an optional comment; a sender with an allowance of at most 32 tokens must leave the comment empty, so at tight budgets discussion is a sequential factual exchange rather than unrestricted debate. Any sender may send an empty packet, so format validity and information transfer are measured separately.

\paragraph{Compact encodings.}
Deterministic senders write the lookup table as a bitstring or as hex digits in a public row order, sending the longest prefix that fits when the query is private and the relevant digit when it is public; one reader decides.

\begin{table*}[htbp]
\centering
\small
\begin{tabular}{p{0.20\textwidth}p{0.65\textwidth}r}
\toprule
Protocol & Information delivered and decision procedure & Calls \\
\midrule
Probability pooling & Local forecasts combined by a mean, summed log odds, or public-evidence-corrected log odds. & $n$ or $n+1$ \\
Fixed factual collection & A deterministic, non-LLM selector sends eligible private facts within the budget; one reader decides. & $1$ \\
Independent selected facts & Each agent selects private fact IDs independently; the reader receives the copied values. & $n+1$ \\
Selected facts + review & Independent selection followed by a second prediction using the same messages and first forecast. & $n+2$ \\
Sequential discussion & Agents select facts in order and see earlier messages; comments follow the per-sender budget rule. & $n+1$ \\
Query/reply & The query holder sends first; the table holder responds with selected facts; the reader decides. & $3$ \\
Complete facts & All private facts are copied to the reader; feasibility is assessed at each delivery budget. & $1$ \\
Compact hex or bitstring & A deterministic encoder sends the table within budget; one reader decides. & $1$ \\
\bottomrule
\end{tabular}
\caption{Protocol access and inference accounting. $n$ is the number of evidence
holders. Corrected pooling includes a public-only forecast. Reused predictions
are charged to every protocol that uses them.}
\label{tab:protocols}
\end{table*}

\subsection{Budgets and models}
Delivery budgets are 64, 256, and 1,024 native model tokens per case, and every copy to a recipient is charged. Senders see their complete-packet costs and a conservative planning guide that limits the number of facts to their allowance divided by the largest one-fact packet cost. Runtime tokenization determines compliance, and oversized or invalid messages are withheld in full.

We evaluate three open-weight models: Qwen3.5-27B and Qwen3.5-9B in pinned Q4\_K\_M quantization, and Gemma-4-E4B-it in QAT Q4\_0 quantization. Each condition uses one generation at temperature zero, after a development-only readiness stage. The two Qwen models form the primary evaluation, one family at two sizes: both run the core suite at all three budgets and the graded sweep. Gemma serves as a cross-family check of which findings depend on the channel and which on the agents (Section~\ref{sec:utilization} and Appendix~\ref{sec:crossfamily}). It runs monitoring, the 60 held-out compatibility cases, and the lookup subset at the 64-token budget, where the protocols differ, plus three lookup runs at temperature 0.7. Two further experiments target utilization and truncation: Qwen 27B runs the compact encodings, together with a query/reply rerun, with a 65,536-token context, and Qwen 9B reruns the lookup subset at reasoning allowances of 4,096 and 8,192 tokens. The original experiments comprise 33,179 task calls, all failures retained. Settings and
provenance are in Appendix~\ref{app:execution}.

\subsection{Outcomes and statistical analysis}
The primary outcome is expected log loss against the exact full-information probability $q=\Pr(Y=1\mid S,U)$. For reported probability $p$, 
\begin{equation}
\ell(p;q)=-q\log\widetilde p-(1-q)\log(1-\widetilde p),
\label{eq:loss}
\end{equation}
where $\widetilde p$ clips $p$ to $[10^{-6},1-10^{-6}]$. This integrates over label noise in monitoring and the graded task. Accuracy, evidence coverage, output validity, and cost are secondary measures. Invalid final answers from the model output are not dropped: they receive a fixed loss of $-\log(10^{-6})\approx13.8$ nats and zero accuracy, a rule set before the evaluation.

The primary comparison is $D(64)$ for Qwen 27B, with fixed collection as the one-shot comparator. Its 95\% interval is a percentile bootstrap (5,000 replicates) over the 100 private/public world pairs, resampled within manual size. Intervals for the compact, reasoning, and Gemma comparisons are exploratory paired-world bootstraps; complete finite censuses, including the graded task, use exact weights without sampled-world intervals.

\subsection{Execution and reproducibility}
\label{app:execution}
\paragraph{Model and dataset.}
The Qwen models are the \texttt{Qwen3.5-9B-GGUF} and
\texttt{Qwen3.5-27B-GGUF} Q4\_K\_M artifacts, and Gemma is \texttt{google/gemma-4-E4B-it-qat-q4\_0-gguf}.

The two lookup sizes each contain 50 independently generated balanced tables with uniform queries; their private/public versions stay together in every worker and bootstrap replicate. The 64 detailed compatibility combinations exhaust the six binary specification fields, each with its original sampled distractor assignment.
The generator and exact evaluator are separate from model input construction.

The final evaluation comprises 10,724 calls per model and 21,448 in total. Readiness and 66 format checks bring the record to 21,850 calls. The audit reconstructs both readiness gates and reproduces all 256 worker scoring files and the combined analyses. Main inference usage is 53,283,520 tokens for Qwen 9B and 45,847,731 for Qwen 27B.

\begin{table}[!htbp]
\centering
\small
\begin{tabular}{lrr}
\toprule
Experiment & Calls & Invalid \\
\midrule
9B reasoning 4,096 & 1,100 & 0 \\
9B reasoning 8,192 & 1,100 & 0 \\
9B graded sweep & 1,152 & 19 \\
27B compact encodings & 2,575 & 5 \\
27B graded sweep & 1,152 & 0 \\
Gemma (all runs) & 4,652 & 70 \\
\midrule
Total & 11,731 & 94 \\
\bottomrule
\end{tabular}
\caption{Model calls and invalid outputs in the additional experiments. Invalid counts include sender packets and probability forecasts that fail validation, not merely incorrect final predictions. The Gemma row combines its temperature-zero replication, three temperature-0.7 lookup runs, and monitoring extension; its 70 invalid outputs comprise 57 sender packets and 13 forecasts, including seven monitoring reader outputs.}
\label{tab:extra-calls}
\end{table}

\subsection{Additional analysis details}
\paragraph{Scoring.}
For monitoring and the graded task, expected Brier score is $(p-q)^2+q(1-q)$, and expected accuracy is $q$ when the selected answer is one and $1-q$ otherwise, averaging over the remaining label noise. Invalid final answers receive Brier penalty one and accuracy zero, and an unavailable probability pool receives the same log-loss penalty as an invalid answer. Infeasible complete-message conditions are labeled separately from prediction failures.

\paragraph{Bootstrap.}
The primary bootstrap samples 50 world pairs with replacement from each manual-size group in every replicate and averages the 100 paired contrasts, using seed 20261105; the 2.5th and 97.5th percentiles of 5,000 replicates form the interval. Qwen 27B's mean private-query gains are 0.381 nats for 16-row manuals and 0.587 for 64-row manuals; Qwen 9B's are 0.381 and 0.849, the latter including its invalid fixed-collection answer. Excluding that answer, Qwen 9B's fixed-collection loss would be 0.482 on the remaining 99 worlds.

\paragraph{Development combinations.}
On the four development compatibility combinations, corrected pooling has mean expected loss 0.803, compared with approximately $10^{-6}$ for fixed collection, for both models.

\paragraph{Message audits.}
Audits compare the fact IDs delivered to the final reader with the evaluator's known sufficient fact sets. A case counts as covered when the reader received the query and the requested row (lookup) or every fact in the sufficient set (compatibility). Because the recorded sufficient sets are not exhaustive, an uncovered case can still be decidable; we therefore report omissions together with the reader's outcome on complete facts. Sender allowances and planning guides are read from the recorded prompts. In the core suite, two Qwen 9B final outputs fail validation, both at the 64-token private-query condition; Qwen 27B has none, and its baseline forecasts match all 1,232 exact reference probabilities (Qwen 9B matches 1,231).

\section{Additional Results}
\label{app:results}

\subsection{All protocols in the three scenarios}
\begin{figure*}[htbp]
\centering
\includegraphics[width=\textwidth]{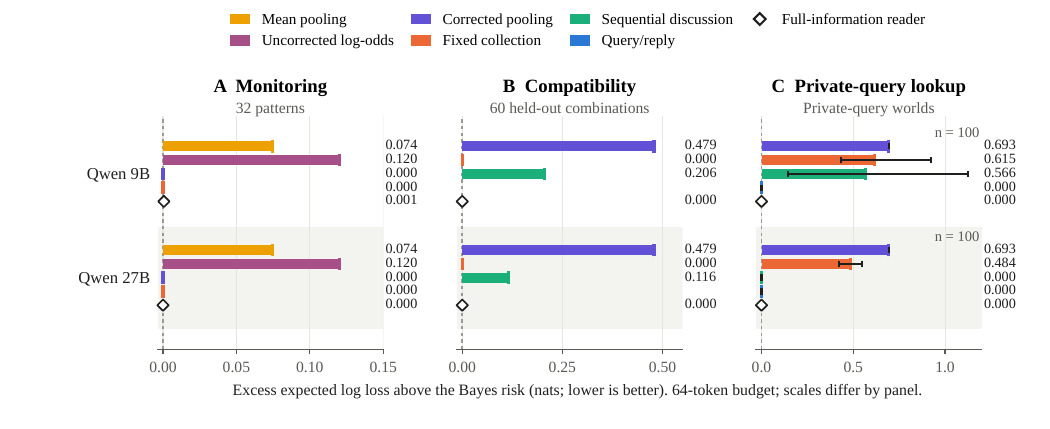}
\caption{Outcomes in the three scenarios: expected log loss above the exact Bayes risk, including invalid outputs (lower is better). Diamonds mark the model's own full-information reader, not the Bayes oracle. Lookup whiskers are exploratory 95\% world-bootstrap intervals; the enumerated tasks have none. Scales differ by panel; protocols are defined in Appendix~\ref{app:protocols}.}
\label{fig:outcome-overview}
\end{figure*}

Figure~\ref{fig:outcome-overview} summarizes the three scenarios at the
64-token delivery limit for the two Qwen models. 

In Figure~\ref{fig:outcome-overview} and Figure~\ref{fig:gemma}, monitoring compares mean, uncorrected log-odds, and corrected pooling with fixed collection; query/reply applies only to lookup. Lookup intervals are 95\% percentile intervals from 5,000 world bootstraps stratified by table size, with the same resamples across protocols. Figure~\ref{fig:gemma} shows Gemma's temperature-zero runs on 32 monitoring patterns, 60 compatibility cases, and 50 lookup worlds, beside Qwen 27B on the same cases; its lookup panel restricts Qwen 27B to the same 50 worlds. Gemma's monitoring losses retain three invalid fixed-collection and four invalid full-information outputs with the fixed penalty, which places them beyond the panel's 0.6-nat cut (2.162 and 2.788 nats above the Bayes risk). Full-information reading has no delivery cap.

Corrected pooling recovers the monitoring posterior within rounding for all 32 reading patterns in both Qwen models, with expected log loss 0.440, equal to fixed and independent factual collection. Mean pooling scores 0.514 and uncorrected log-odds pooling 0.560. With an uninformative public inspection, uncorrected and corrected pooling both score 0.482; at reliability 0.75 they score 0.638 and 0.398. Removing the repeated public contribution restores the loss predicted by Equation~\ref{eq:pool}.

Compatibility favors exchanging the underlying attributes (Figure~\ref{fig:outcome-overview}; Appendix Table~\ref{tab:compatibility}). On the 60 held-out combinations, corrected pooling answers 44 correctly for each Qwen model, whereas fixed and independent factual collection answer all 60. On the four simple interface patterns, every truthful local posterior is one-half, so all three pools reach 2/4 and factual collection 4/4. Because both models report the exact local posteriors, the pooling result measures the channel, not the model.

\begin{table}[htbp]
\centering
\small
\begin{tabular}{lrr}
\toprule
& Qwen 9B & Qwen 27B \\
Method & Correct / loss & Correct / loss \\
\midrule
Corrected pooling & 44 / 0.479 & 44 / 0.479 \\
Fixed collection & 60 / 0.000 & 60 / 0.000 \\
Selected facts & 60 / 0.010 & 60 / 0.000 \\
Discussion & 50 / 0.206 & 53 / 0.116 \\
Complete facts (reference) & 60 / 0.000 & 60 / 0.000 \\
\bottomrule
\end{tabular}
\caption{Detailed compatibility: 60 held-out combinations at a 64-token
delivery limit. Loss is expected log loss in nats; 0.000 denotes approximately
$10^{-6}$ under clipping.}
\label{tab:compatibility}
\end{table}

\subsection{Graded information loss}
\label{app:graded-results}

\begin{figure*}[htbp]
\centering
\includegraphics[width=\textwidth]{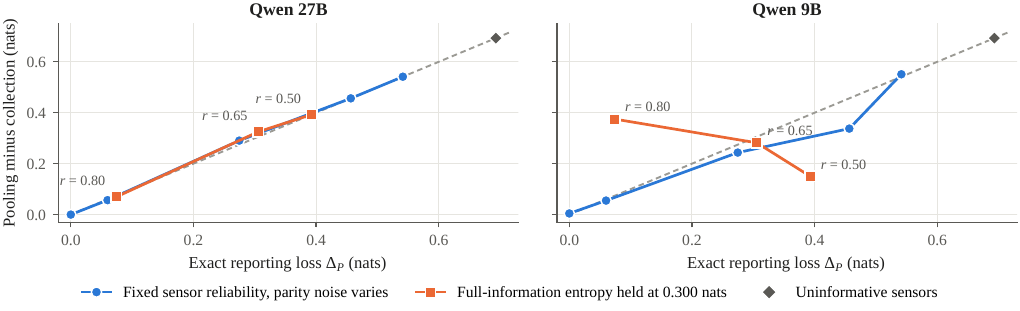}
\caption{Graded information loss. Each point is an exact weighted census of 32
observations: the exact reporting loss $\Delta_P$ against the measured loss
gap between corrected pooling and complete factual collection (positive
favors collection). The dashed line is the oracle identity, not a fit; labels
give sensor reliability in the matched-entropy series. Qwen 27B has no invalid
outputs. Qwen 9B gaps use the 269 of 288 cases with a valid reader output;
gaps that count invalid answers are in Table~\ref{tab:graded}.}
\label{fig:mixed}
\end{figure*}

Figure~\ref{fig:mixed} compares the exact reporting loss $\Delta_P$
with the observed pooling-minus-collection loss across nine
settings. For Qwen 27B, the discrepancy is at most 0.020 nats
in every setting, including those with full-information Bayes
risk fixed at 0.300 nats. Qwen 9B deviates more substantially;
Appendix~\ref{app:graded-9b} examines its probability-estimation
and full-information reader errors.

\subsection{Selection: indexed messages across budgets}
\begin{figure}[htbp]
\centering
\includegraphics[width=0.48\columnwidth]{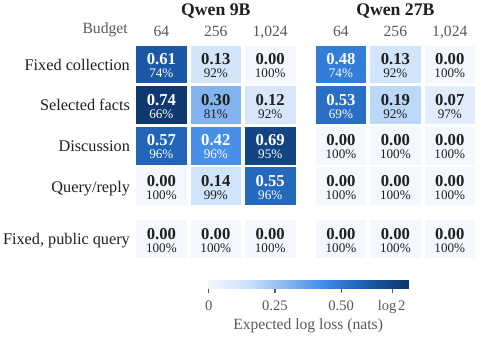}
\caption{Private-query lookup with indexed messages over the same 100 worlds, including invalid answers. Columns are delivery budgets; each cell gives expected log loss (shading, bold) and accuracy. One-shot protocols improve as more rows fit; Qwen 9B's sequential protocols degrade at larger budgets because of omitted queries (Section~\ref{sec:audits}). The bottom row is the public-query control.}
\label{fig:lookup}
\end{figure}

At 64 delivered tokens, query/reply answers all 100 private-query lookup cases for both Qwen models, compared with 74 for fixed collection (Figure~\ref{fig:lookup}). Independent selected facts reach 66 and 69, and sequential discussion 96 and 100, for Qwen 9B and Qwen 27B. Every probability pool scores $\log 2\approx0.693$ nats, because each local posterior is one-half (Table~\ref{tab:lookup}). For the primary model, the paired private-minus-public contrast is $D(64)=0.484$ nats (95\% CI [0.421, 0.548]); for Qwen 9B it is 0.615 nats ([0.435, 0.922]), including one invalid fixed-collection answer. With a public query, fixed collection and query/reply both answer 100/100 at every budget, and fixed collection sends only the requested row, at 10 tokens.

The budget sweep supplies a second boundary. A complete indexed packet costs 124 tokens for a 16-row manual and 460 for a 64-row manual; at the 1,024-token limit, fixed collection sends the whole table and answers 100/100 for both models. In the indexed format, the advantage of asking first depends jointly on query visibility and on the cost of transmitting the table.

\begin{table}[htbp]
\centering
\small
\begin{tabular}{lrrrr}
\toprule
& \multicolumn{2}{c}{Qwen 9B} & \multicolumn{2}{c}{Qwen 27B} \\
\cmidrule(lr){2-3}\cmidrule(lr){4-5}
Protocol & Corr. & Loss & Corr. & Loss \\
\midrule
Corrected pooling & 52 & 0.693 & 52 & 0.693 \\
Fixed collection & 74 & 0.615 & 74 & 0.484 \\
Selected facts & 66 & 0.740 & 69 & 0.535 \\
Selected + review & 65 & 0.615 & 69 & 0.535 \\
Discussion & 96 & 0.566 & 100 & 0.000 \\
Query/reply & 100 & 0.000 & 100 & 0.000 \\
\bottomrule
\end{tabular}
\caption{Private-query lookup at the 64-token delivery limit over all 100 worlds (correct of 100; expected log loss). Invalid final answers receive the fixed loss; 0.000 denotes approximately $10^{-6}$.}
\label{tab:lookup}
\end{table}

\subsection{Compact encodings}
\label{sec:compact}
\begin{figure*}[htbp]
\centering
\includegraphics[width=\textwidth]{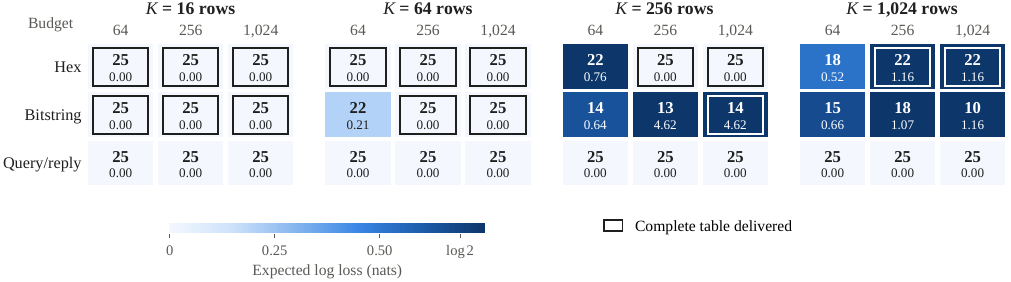}
\caption{Qwen 27B compact-encoding control, private query, with a 4,096-token reasoning allowance. Each panel is one table size with 25 worlds; bold numbers count correct answers and small numbers give expected log loss, which also sets the shading. Outlined cells delivered the complete table. Hex matches query/reply on small tables. At 1,024 rows the complete hex table fits within 256 delivered tokens and yields 22 correct answers in 25 cases. Identical prompts reused across budgets are the same prediction.}
\label{fig:compact}
\end{figure*}

A compact encoding changes the result at the 16- and 64-row sizes (Figure~\ref{fig:compact}). Complete hex messages, including the private query, cost 19--21 tokens at $K=16$ and 28--33 at $K=64$, and at the 64-token budget hex collection and query/reply both answer all 50 sampled worlds. The indexed-format advantage disappears under this representation.

Larger tables separate delivery from utilization. At budget 64, hex answers 22/25 for $K=256$ and 18/25 for $K=1024$, while query/reply answers 25/25 in both. Complete 256-row hex packets cost 66--78 tokens, and at budget 256 hex reaches 25/25. Complete 1,024-row packets cost 233--251 tokens and also fit at budget 256, yet the reader answers only 22/25: two confident errors and one incorrect tie give mean loss 1.161, against approximately zero for query/reply ($D=1.161$ nats, exploratory 95\% CI [0.028, 2.819]). With 25 worlds the interval is wide, but full disclosure and successful decoding are clearly distinct conditions. The bitstring control, which is longer, answers 25, 22, 14, and 15 of 25 at budget 64 across table sizes, against hex's 25, 25, 22, and 18: dense encodings can make reading harder as well as cheaper.

\paragraph{Matched reasoning-allowance rerun.}
We reran the Qwen 27B reader on the same 25 private-query worlds at $K=1024$ with the 256-token delivery budget. The reader received the complete hex table and query in every case (233--251 delivered tokens), and the prompts were byte-identical across runs. With the original 4,096-token reasoning allowance, the reader answered 22/25 correctly. With a 32,768-token allowance, it answered 25/25. Both runs used temperature zero and seed 42. The original run used a 65,536-token context window and the rerun used 81,920 tokens. Both windows fit the prompts and the full generation allowance of either run.

\subsection{Message audits}
\label{sec:audits}
In the core suite, all 4,992 sender outputs per Qwen model satisfy format, ownership, and budget rules. Auditing their contents locates the remaining deficits in what was sent.

\paragraph{Discussion errors accompany omitted attributes.}
Every one of the 10 Qwen 9B and 7 Qwen 27B compatibility errors in discussion lacks a fact from the recorded sufficient set, and the same model answers each case correctly when given complete facts. The omissions follow from how the sequential protocol divides the 64-token limit: the first speaker receives a 21-token allowance and a planning guide of at most two facts, although its three relevant facts cost 19 tokens. The order's temperature requirement is missing from 46 of 60 Qwen 9B and 40 of 60 Qwen 27B transcripts. Independent selection, whose senders each receive 32 tokens and a four-fact guide, answers all 60. Budget allocation is part of the protocol comparison, not evidence that exchange itself degrades decisions.

\paragraph{Model senders send fewer rows than fit.}
In private-query lookup at 64 tokens, the deterministic collector sends seven rows, including the requested row in 29 of 100 cases. LLM senders, guided to select at most five rows, transmit 2.8 rows on average for Qwen 9B and 4.9 for Qwen 27B; the requested row reaches the reader in only 10 and 19 cases, respectively. Because every row is equally likely to be queried, these shorter messages reduce the chance of delivering the needed information and contribute to the lower accuracy of LLM-selected one-shot communication.

\paragraph{Missing queries and confident guesses.}
Qwen 27B delivers the query and requested row in all 100 query/reply and discussion cases at every budget. Qwen 9B's query holder sends an empty but valid packet in 1, 4, and 6 of 100 query/reply cases at 64, 256, and 1,024 tokens, and in 7, 6, and 8 discussion cases. All 18 Qwen 9B errors in the two sequential protocols occur where the query never reached the reader, and in 17 of them the reader assigned probability 0 or 1. This is the rising loss in Figure~\ref{fig:lookup}: accuracy stays at 94--100\%, but each missing query can become a confident error. Extra reader review changes 4 of 900 Qwen 9B probabilities and none of Qwen 27B's.

\subsection{Delivered tokens versus inference cost}
\begin{figure*}[htbp]
\centering
\includegraphics[width=.9\textwidth]{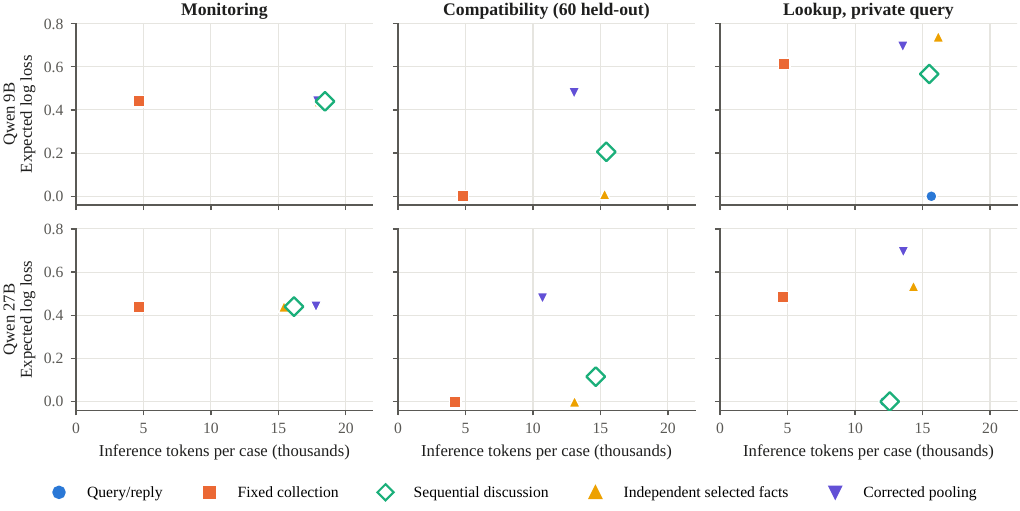}
\caption{Expected log loss against mean inference tokens per case at the 64-token delivery limit (lower left is better).}
\label{fig:cost}
\end{figure*}

The protocol that minimizes delivered tokens need not minimize inference (Appendix Figure~\ref{fig:cost}). In private-query lookup at 64 tokens, query/reply delivers 28 tokens against 61 for fixed collection but needs three model calls instead of one: mean inference is 15,654 versus 4,740 tokens for Qwen 9B and 12,203 versus 4,680 for Qwen 27B. In monitoring, corrected probability reports and fixed collection reach the same loss, but the three readings cost 24 delivered tokens against 33.6 for the reports, and collection needs one inference call rather than four forecasts. Encoding changes the trade-off. At $K=16$, hex delivers 20.4 tokens, fewer than query/reply's 28, and both are correct. At $K=1024$ and budget 256, full hex uses about 5,095 inference and 243.6 delivered tokens per case, while query/reply uses 42,604 inference and 28 delivered tokens: its higher accuracy costs about eight times the inference.

\section{Robustness and Second Model Family}
\label{app:robust}
\label{app:sensitivity}

\begin{figure*}[htbp]
\centering
\includegraphics[width=\textwidth]{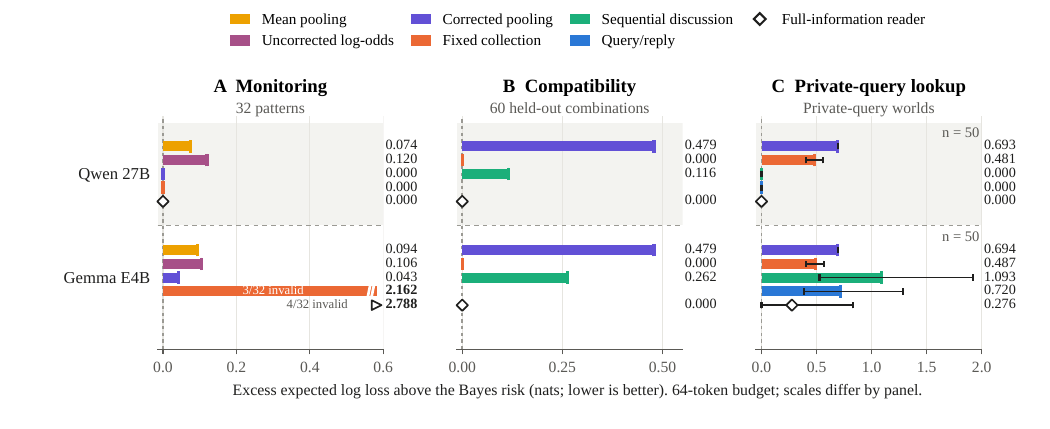}
\caption{Cross-family check at the 64-token budget: Qwen 27B and Gemma E4B on the same cases, in the layout of Figure~\ref{fig:outcome-overview}.}
\label{fig:gemma}
\end{figure*}

\subsection{Qwen 9B in the graded sweep}
\label{app:graded-9b}
The Qwen 9B graded sweep illustrates how errors in probability estimation and full-information prediction can obscure the theoretical reporting loss. Its full-information reader produces 19 invalid predictions among 288 cases; including their penalties, pooling has lower loss than factual collection in six of nine settings (Table~\ref{tab:graded}). Among the 269 cases with valid reader outputs, the pooling penalty generally increases with $\Delta_P$ in the fixed-reliability series, with gaps of 0.005, 0.055, 0.244, 0.338, and 0.551 nats.

The matched-entropy series behaves differently: as $\Delta_P$ decreases from 0.393 to 0.074 nats, the observed gap increases from 0.150 to 0.375. Two sources of model error contribute to this reversal. Local probability reports deviate increasingly from their exact values as sensor reliability rises (mean absolute errors of 0.000, 0.084, and 0.187), while the full-information reader incurs losses of 0.455--0.543 nats despite a fixed Bayes risk of 0.300. These deviations reflect imperfect model execution rather than changes to the theoretical reporting loss.

\paragraph{Graded distributions.}
The state has prior probability one-half. Conditional on the state, sensor errors, shared-inspection noise, and parity noise are independent. Code pairs are uniform within each parity, so either code bit alone is uninformative. An uninformative shared inspection (reliability 0.50) supplies the fifth observed binary value. The sensor reliability is $r=0.70$ for parity-noise levels $\eta\in\{0.50,0.30,0.10,0.02,0\}$; the separate endpoint uses $r=0.50$ and $\eta=0$. Matched-entropy settings use $r\in\{0.50,0.65,0.80\}$ and $\eta=0.0889062695$, $0.1039943074$, and $0.2291892038$, respectively, solved before inference to give full-information entropy 0.300 nats. Both Qwen models report local probabilities and read all factual records without a delivery cap. The evaluator integrates over the conditional label probability; stored placeholder labels are not scored as observed outcomes. Reasoning caps are reached in 1,144/1,152 graded 9B calls and all 1,152 27B calls; the larger-allowance lookup check does not cover this task.

\paragraph{Graded sensitivities.}
Table~\ref{tab:graded} reports the gaps with invalid answers counted beside the valid-output gaps plotted in Figure~\ref{fig:mixed}. All 19 invalid 9B predictions come from the full-information reader: 16 answer/probability disagreements, two precision violations, and one malformed vector. A post-hoc probability-vector sensitivity, which scores well-formed vectors
regardless of answer-field agreement or decimal precision, gives
fixed-reliability gaps of 0.005, 0.059, 0.232, $-0.504$, and 0.199 nats; the one malformed vector keeps its penalty. Valid-output summaries condition on selected outputs and are reported separately from the complete-population comparison.
\begin{table}[htbp]
\centering
\small
\begin{tabular}{lrrrr}
\toprule
& & 27B & \multicolumn{2}{c}{9B gap} \\
\cmidrule(lr){4-5}
Setting & $\Delta_P$ & gap & incl. & valid \\
\midrule
$r=0.70$, $\eta=0.50$ & 0.000 & 0.000 & 0.005 & 0.005 \\
$r=0.70$, $\eta=0.30$ & 0.060 & 0.057 & $-0.630$ & 0.055 \\
$r=0.70$, $\eta=0.10$ & 0.275 & 0.290 & $-0.112$ & 0.244 \\
$r=0.70$, $\eta=0.02$ & 0.457 & 0.457 & $-1.162$ & 0.338 \\
$r=0.70$, $\eta=0$ & 0.541 & 0.541 & 0.051 & 0.551 \\
Matched, $r=0.80$ & 0.074 & 0.069 & $-1.342$ & 0.375 \\
Matched, $r=0.65$ & 0.306 & 0.325 & $-1.282$ & 0.282 \\
Matched, $r=0.50$ & 0.393 & 0.393 & $-0.680$ & 0.150 \\
Endpoint & 0.693 & 0.693 & 0.693 & 0.693 \\
\bottomrule
\end{tabular}
\caption{Graded sweep: exact reporting loss and pooling-minus-collection gaps in nats. ``incl.'' counts invalid answers at the fixed loss; ``valid'' uses the 269 of 288 cases with a valid 9B reader output. Qwen 27B has no invalid outputs.}
\label{tab:graded}
\end{table} 

\subsection{Reasoning allowance}
Qwen 9B reached its reasoning allowance in most calls. Doubling the allowance from 4,096 to 8,192 tokens reduced cap hits from 818/1,100 (74.4\%) to 397/1,100 (36.1\%) of task calls but left fixed collection (40/50 private-query cases correct) and query/reply (50/50) unchanged, so $D(64)=0.481$ nats (95\% CI [0.401, 0.557]) in both arms. Discussion did not improve: missing queries rose from 2 to 5 of 50, and its private-query loss from 0.318 to 0.871 nats. Truncation explains neither the contrast nor the omissions.

\subsection{Cross-family comparison: information limits and model execution}
\label{sec:crossfamily}

We evaluate Gemma to examine which findings extend beyond the Qwen models (Figure~\ref{fig:gemma}). The theoretical information limits are model-independent, but observed performance also depends on how accurately agents produce probability reports, select evidence, and use received information.

\paragraph{Probability-reporting results.}
In compatibility, Gemma's corrected pooling answers 44/60 cases (loss 0.479), compared with 60/60 under fixed factual collection, matching both Qwen models. In monitoring, corrected pooling also outperforms mean pooling and uncorrected log-odds pooling (losses of 0.483, 0.534, and 0.546, respectively), although it remains 0.043 nats above the Bayes risk.

The residual errors reflect imperfect model execution. All 128 local and public probability reports are valid, but eight of the 96 local reports differ from their exact Bayesian values (weighted mean absolute error 0.027). Moreover, fixed collection delivers all three monitoring readings, yet three reader predictions are invalid, as are four full-information reader predictions. Thus, the qualitative pooling results extend to Gemma, while accurate reporting and utilization remain model-dependent.

\paragraph{Query-first performance depends on model execution.}
In private-query lookup at temperature zero, Gemma answers 37/50 cases under fixed collection (loss 0.487), compared with 35/50 under query/reply (loss 0.720), giving $D(64)=-0.192$ nats (95\% CI [$-0.758$, 0.143]). Message audits reveal that the query is missing in 32 of 50 exchanges, including 31 valid but empty packets. In all 16 cases where both the query and requested row reach the reader, the prediction is correct.

Three additional runs at temperature 0.7 show variable performance, with a mean $D(64)$ of 0.319 nats across seeds (95\% CI [$-0.050$, 0.785]). These results contrast with Qwen 27B's reliable query delivery and show that the empirical benefit of asking first depends on successful protocol execution.

\begin{table}[htbp]
\centering
\small
\begin{tabular}{lrrrr}
\toprule
& \multicolumn{2}{c}{Fixed collection} & \multicolumn{2}{c}{Query/reply} \\
\cmidrule(lr){2-3}\cmidrule(lr){4-5}
Seed & Correct & Loss & Correct & Loss \\
\midrule
101 & 39/50 & 0.859 & 33/50 & 0.734 \\
202 & 38/50 & 1.004 & 38/50 & 0.363 \\
303 & 40/50 & 0.481 & 36/50 & 0.457 \\
\bottomrule
\end{tabular}
\caption{Gemma private-query lookup at temperature 0.7 and budget 64 on the same 50 manuals. Scores include invalid predictions. These are repeated generations on 50 worlds, not 150 independent worlds.}
\label{tab:gemma-seeds}
\end{table}

\noindent\textbf{Gemma across generation seeds.}
Table~\ref{tab:gemma-seeds} reports the three temperature-0.7 runs. Query/reply achieves lower expected log loss in each run but does not improve accuracy. 
Restricting comparisons to cases where both protocols produce valid predictions gives fixed-minus-query loss differences of $-0.140$, $0.121$, and $0.023$ nats, indicating that invalid fixed-collection outputs contribute to the aggregate loss advantage. 
Queries are omitted in 33, 26, and 35 exchanges, respectively. 
These results further illustrate the importance of auditing delivered information alongside final predictions.

Gemma's compatibility cases were evaluated once, at temperature zero; the seed runs cover lookup only. 
In this run, independently selected facts answer 59/60 cases and sequential discussion answers 49/60.

\subsection{Paired uncertainty}
Exploratory intervals for the compact, reasoning, and Gemma comparisons use 5,000 bootstrap draws. We resample entire manuals with their private/public versions and protocol outcomes. Size-aggregated lookup comparisons resample within table size; compact panels resample their 25 manuals. Budget aliases do not constitute independent observations. 

\end{document}